\documentclass[runningheads]{llncs}
\usepackage{graphicx}
\usepackage{colortbl}
\usepackage{amsmath}
\usepackage{amsfonts}
\usepackage[ruled,vlined,linesnumbered]{algorithm2e}
\usepackage{lineno}
\usepackage{pifont}
\usepackage{url} 
\usepackage{bm}
\usepackage{subcaption}
\usepackage[T1]{fontenc}

\begin{document}
\title{MTGP: Combining Metamorphic Testing and Genetic Programming}
\author{Dominik Sobania \and Martin Briesch \and Philipp Röchner \and Franz Rothlauf}
\authorrunning{Sobania et al.}
\institute{Johannes Gutenberg University, Mainz, Germany\\
\email{\{dsobania,briesch,proechne,rothlauf\}@uni-mainz.de}}

\maketitle             
\begin{abstract}
Genetic programming is an evolutionary approach known for its performance in program synthesis. However, it is not yet mature enough for a practical use in real-world software development, since usually many training cases are required to generate programs that generalize to unseen test cases. As in practice, the training cases have to be expensively hand-labeled by the user, we need an approach to check the program behavior with a lower number of training cases. Metamorphic testing needs no labeled input/output examples. Instead, the program is executed multiple times, first on a given (randomly generated) input, followed by related inputs to check whether certain user-defined relations between the observed outputs hold. In this work, we suggest MTGP, which combines metamorphic testing and genetic programming and study its performance and the generalizability of the generated programs. Further, we analyze how the generalizability depends on the number of given labeled training cases. We find that using metamorphic testing combined with labeled training cases leads to a higher generalization rate than the use of labeled training cases alone in almost all studied configurations. Consequently, we recommend researchers to use metamorphic testing in their systems if the labeling of the training data is expensive.  
\end{abstract}
\section{Introduction}

Genetic programming (GP)~\cite{cramer1985representation,koza1992genetic} is an evolutionary algorithm-based approach to automatically generate programs in a given programming language that meet user-defined requirements. In GP-based program synthesis, these specifications are usually given as input/output examples, which define the expected output from a generated program for a given input, and are used as training data during the evolutionary search. 

With the introduction of new selection methods~\cite{helmuth2020importance,spector2012assessment}, variation operators~\cite{helmuth2018program} and grammar design techniques~\cite{forstenlechner2017grammar}, GP-based program synthesis has made great progress in the last years. Recently, it has been shown that GP-based approaches for program synthesis are even competitive in performance with state-of-the-art neural network-based approaches~\cite{sobania2022choose}. 

Unfortunately, GP-based program synthesis is not yet mature enough for a practical use in real-world software development, since many input/output examples are usually required (up to $200$ are regularly used in the literature \cite{helmuth2021psb2}) to generate programs that not only work on the training cases, but also generalize to previously unseen test cases. In practice, however, the training cases have to be labeled manually by the user which is very expensive and time consuming. So it is necessary to reduce the number of required training cases in order to minimize the user's manual effort. However, simply reducing the number of training cases is not sufficient, since a small training set can easily be overfitted which will on average lead to a poor generalization of the generated programs. Consequently, we need a supplementary approach to specify and check the desired program behavior without adding more manually defined training cases.

With metamorphic testing~\cite{chen1998metamorphic}, we do not need additional hand-labeled training cases as we execute a program multiple times (starting with a random input, followed by related inputs) and check whether the relations between the observed outputs logically fit the user's domain knowledge. E.g., a function that assigns a grade based on the score achieved in an exam could be executed with a random score. If we then increase this score, the function must return an equal or better grade, otherwise the metamorphic relation is violated and the function must be incorrect. Such metamorphic relations, where labeling is not necessary, could be used together with a classical (but smaller) hand-labeled training set. We expect that these additional relations help to improve the generalization ability of GP-generated programs.

Therefore, in this work, we suggest an approach that combines metamorphic testing and genetic programming (MTGP) and study its performance and the generalization ability of the generated programs on a set of common program synthesis benchmark problems. To analyze how GP's generalization ability depends on the number of given training cases, we perform experiments for different (labeled) training set sizes. 

MTGP is based on a grammar-guided GP approach which uses lexicase~\cite{spector2012assessment} for the selection of individuals during evolution. Since lexicase selection is not based on an aggregated fitness value, but considers the performance on individual cases, it is well suited to take hand-labeled training cases in combination with metamorphic relations into account during selection. To study this combination, we analyze different sizes of the hand-labeled training set and add further tests based on the metamorphic relations defined for the considered benchmark problem. More specific, the tests based on the metamorphic relations can be constructed automatically based on random inputs. A candidate program is executed first on the random input and then on a follow-up input (based on the random input). After that, the outputs of the candidate program on the random input and the follow-up input are compared regarding to a pre-defined metamorphic relation. If the relation holds, the test is passed, otherwise it is failed. So the outcome of a metamorphic test can therefore be treated in the same way as that of a test based on labeled input/output examples, with the difference that no expensive manual labeling is necessary for an automatically generated metamorphic test case. In our experiments, we find that incorporating metamorphic testing in combination with hand-labeled training cases leads to a higher generalization rate than the use of hand-labeled training cases alone (as usual in GP-based program synthesis) in almost all studied configurations.

Following this introduction, we present in Sect.~\ref{sec:relatedwork} recent work related to GP-based program synthesis and work on metamorphic testing. In Sect.~\ref{sec:method} we describe metamorphic testing as well as its integration into GP in detail. Furthermore, we present the used program synthesis benchmark problems together with their associated metamorphic relations. In Sect.~\ref{sec:experiments} we describe our experimental setup and discuss the results before concluding the paper in Sect.~\ref{sec:conclusion}.

\section{Related Work}\label{sec:relatedwork}

The main approaches in GP-based program synthesis are stack-based GP and grammar-guided GP \cite{sobania2022comprehensive}. These approaches differ primarily in their program representation and their techniques used to support different data types. 

Stack-based GP approaches use different stacks for the separation of different data types \cite{spector2002genetic}. During program execution, data is taken as input from the appropriate stacks and the results are pushed back to the associated stack. In current systems, the individual program instructions are also on their own stack, which allows changes to the program flow at runtime \cite{spector2005push3}. 

Grammar-guided GP approaches use a context-free grammar to represent the supported control structures and statements in their relationship to each other and to distinguish different data types \cite{forstenlechner2016grammar,whigham1995grammatically}. In principle, this technique can be used to create programs in any programming language. In recent years, however, using grammar-guided GP approaches, mainly Python programs have been evolved \cite{forstenlechner2017grammar,schweim2022effects,sobania2020challenges}.  

Regardless of the GP approach used, the program synthesis results have been significantly improved in recent years, primarily through the use of lexicase selection and its variants \cite{forstenlechner2017grammar,helmuth2020benchmarking,helmuth2016lexicase,helmuth2020importance,helmuth2020explaining}. In contrast to selection methods such as tournament selection, lexicase selection is not based on an aggregated fitness value (see evaluation bottleneck \cite{krawiec2016behavioral}), but selects on the basis of the results on individual training cases \cite{spector2012assessment} which allows to include also the structure of the given training data. 

Also independent of the used approach, mainly input/output examples are used for training in GP-based program synthesis. However, there exists also work which uses additional information, such as the textual description of the problem \cite{hemberg2019domain} or formal constraints \cite{blkadek2018counterexample} to improve the program synthesis performance.

Since the input/output examples given as training data are only an incomplete problem definition, it is important that the generated programs not only work on the training data, but also produce correct results on previously unseen inputs. In order to improve the generalization ability of the programs generated by GP, the literature knows approaches that generate smaller programs, either by post-simplification \cite{helmuth2017improving} or by a selection at the end of a run \cite{sobania2021generalizability}. Another option is to use batch lexicase selection \cite{aenugu2019lexicase} to improve generalizability \cite{sobania2022program}. In addition to that, recently a method has been presented that can be used to predict whether the programs generated by GP will generalize to unseen data or not \cite{sobania2021generalizability2}. 

Metamorphic testing introduced by Chen et al.~\cite{chen1998metamorphic} is a method from software development that allows to check certain properties in the program under test without the need of explicitly specifying the expected output of a test (see Sect.~\ref{sec:metamorphictesting} for a detailed description). In the field of evolutionary computation, metamorphic testing was used, e.g., for the genetic improvement of existing software \cite{langdon2020evolving}. 

However, to the best of our knowledge, no work so far studied the impact of combining metamorphic testing and GP on the program synthesis performance and the generalizability of the generated programs. 

\section{Methodology}\label{sec:method}

In this section, we describe the basics of metamorphic testing and show how it works with some illustrative examples. Furthermore, we present the selected program synthesis benchmark problems together with their metamorphic relations. Lastly, we describe in detail how metamorphic testing and GP-based program synthesis can be combined.

\subsection{Metamorphic Testing}\label{sec:metamorphictesting}

Metamorphic testing~\cite{chen1998metamorphic} is a method from software development to check if certain logic properties hold in a given function $f$. These properties are defined by so-called metamorphic relations which describe the logic connection between a given (random) base input $I$ with its observed output $f(I)$ and a further follow-up input $I'$ with its corresponding output $f(I')$.\footnote{More than one follow-up test is also possible, but in this work we focus on exactly one follow-up test.} The key advantage of metamorphic testing compared to classical test methods is that we need no expensive labeled input/output examples as we are just interested if the metamorphic relation between the observed outputs $f(I)$ and $f(I')$ holds.

\begin{figure}[ht!]
\centering
   \begin{subfigure}[b]{0.94\textwidth}
   \includegraphics[width=1.0\textwidth]{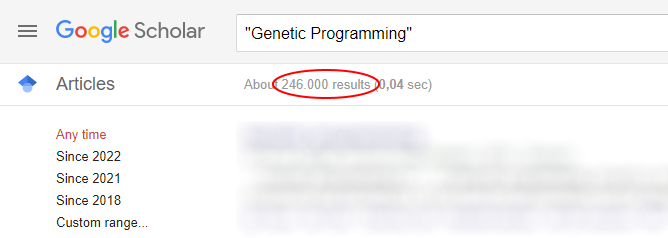}
   \caption{Search result for base input $I$ (no filter).}
   \label{fig:Ng1} 
\end{subfigure}\vspace{0.2cm}
\\
\begin{subfigure}[b]{0.94\textwidth}
   \includegraphics[width=1.0\textwidth]{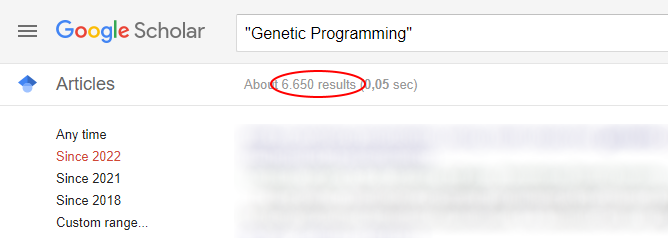}
   \caption{Search result for follow-up input $I'$ (active filter).}
   \label{fig:Ng2}
\end{subfigure}
\caption{Screenshots from Google Scholar illustrating a simple example of metamorphic testing. For the base input $I$ (a), we see the corresponding output $f(I) = 246,000$ and for the follow-up input $I'$ (b) the output is $f(I') = 6,650$, so the defined relation $f(I) \geq f(I')$ holds.}\label{fig:googleexample}
\end{figure}

As a first intuitive example (based on the example given in \cite{arrieta2022multi}), we can think of a simple web search. E.g., as base input $I$ we search for the exact term "Genetic Programming" without filters in a scientific search engine and find $f(I)$ results. As follow-up input $I'$ we search for the same term "Genetic Programming" but limit the results to publications since $2022$ and get $f(I')$ results. As metamorphic relation, we define that $f(I) \geq f(I')$, as additional filters should lead to an equal or lower number of results. Figure~\ref{fig:googleexample}, shows screenshots from Google Scholar illustrating this example. We see that $f(I) = 246,000$ and $f(I') = 6,650$, so the metamorphic relation holds as $f(I) \geq f(I')$.

\begin{figure}[ht!]
  \centering
  \includegraphics[width=0.7\textwidth]{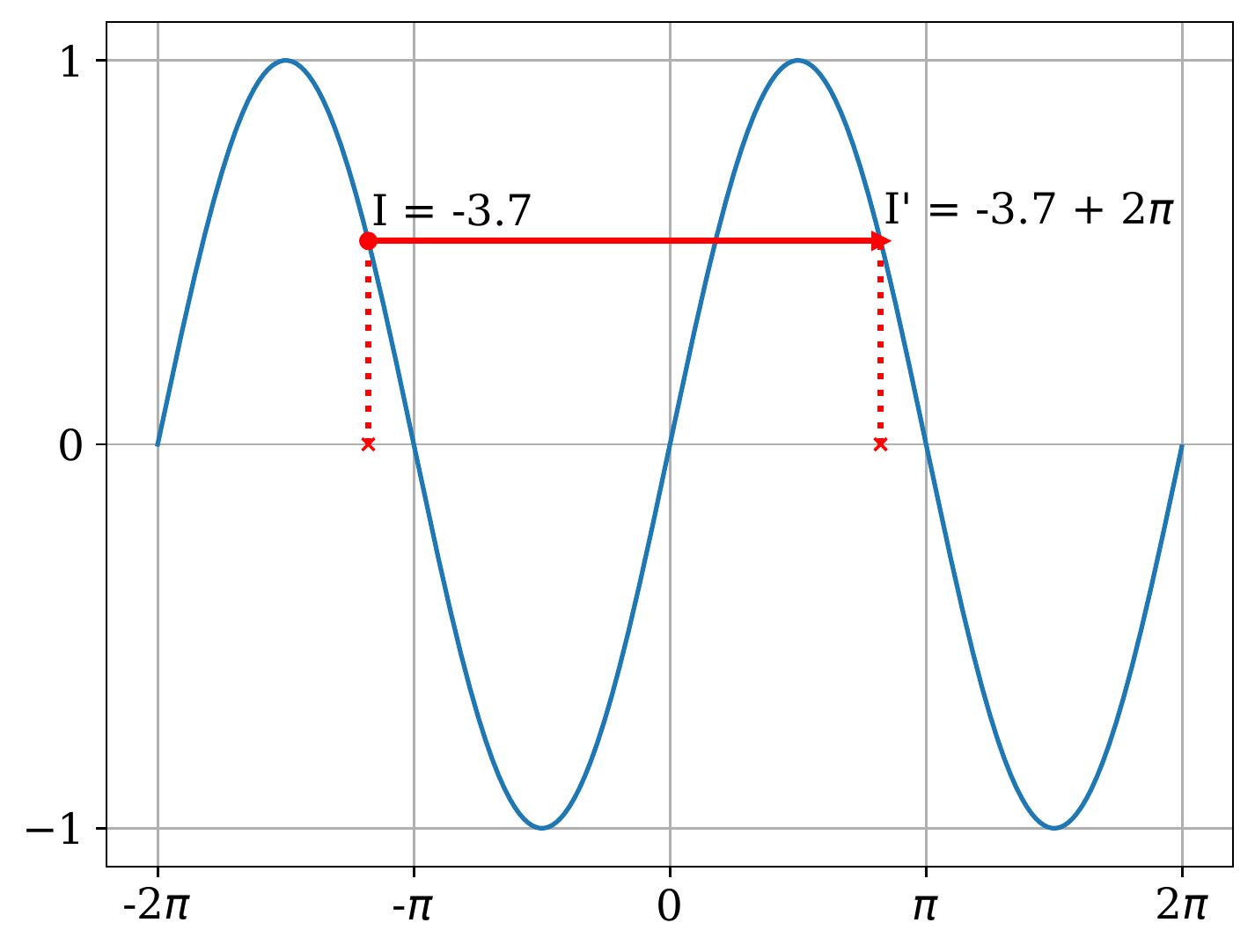}
  \caption{An example of the sine function. We see that the result of the base input $I = -3.7$ and the follow-up input $I' = -3.7 + 2\pi$ is identical, so the defined metamorphic relation $f(I) = f(I')$ is satisfied.}
  \label{fig:sineexample}
\end{figure}

As a second, more technical, example, we choose the sine function, where we expect that it is $2\pi$ periodic. Consequently, for the inputs $I$ and $I' = I + 2\pi$, a metamorphic relation could be defined as $f(I) = f(I')$ \cite{chen2018metamorphic,chen2004metamorphic}. Figure~\ref{fig:sineexample} illustrates this example for $I = -3.7$. We see that constructed follow-up input $I' = -3.7 + 2\pi$ leads to the same result, so the relation $f(I) = f(I')$ holds. 

\subsection{Benchmark Problems and Metamorphic Relations}

For the evaluation, we selected three problems from the program synthesis benchmark suite by Helmuth and Spector~\cite{helmuth2015general} which differ both in difficulty, according to a recent meta study~\cite{sobania2022comprehensive}, as well as in the data types used for input and output. Furthermore, we define two metamorphic relations for each of the benchmark problems for better comparison. However, more metamorphic relations are conceivable. We focused on relatively simple metamorphic relations which can be formulated by a user even with basic domain knowledge about the problem. The benchmark problems and metamorphic relations are defined as follows: 
\\\\
\\\\
\noindent \textbf{Count Odds Problem:}
\\\\
\noindent \textbf{Definition}: A program should be generated that returns the number of odd values in a given list of integers. More formal, we search for a function $f$ that maps the given vector of integers $I = (x_1, \dots, x_n) \in \mathbb{Z}^n$ to the number of odds $f(I) = c \in \mathbb{N}_0$ as return value. 
\\\\ 
\noindent \textbf{Metamorphic Relations}: \textbf{\textit{1)}} As first metamorphic relation, we require that the output of the program under test does not change when the input list is extended by an arbitrary number of even integers. For a given input $I = (x_1, \dots, x_n)$ and the program $f$ we calculate $f(I) = c_I$. As follow-up test, we create an extended input $I' = (x_1, \dots, x_n, 2, 4, 6, \dots, 2, 4, 6)$ by extending the input $I$ with a random repetition of the vector $(2, 4, 6)$. The corresponding output of the manipulated input is $f(I') = c_{I'}$. This relation holds if $c_I = c_{I'}$. We chose 2, 4, and 6 as we expect that these values are known to be even numbers for users with basic domain knowledge about the problem (no knowledge about the \texttt{modulo} function needed). \textbf{\textit{2)}} Analogously, we require for a second metamorphic relation that the output value increases if we extend the input by an arbitrary number of odd integers. Contrarily to the first relation, we extend the input $I' = (x_1, \dots, x_n, 1,3,5,\dots,1,3,5)$ by a random number of repetitions of the vector $(1,3,5)$. Consequently, the corresponding output is $f(I') = c_{I'}$. This relation holds if $c_I < c_{I'}$. This time, we chose 1, 3, and 5 as trivial odd numbers. 

\clearpage

\noindent \textbf{Grade Problem:}
\\\\
\noindent \textbf{Definition}: We search a program that maps the numeric score achieved by a student to a discrete grade based on given thresholds. 
Specifically, we search for a function $f$ that maps the input $I$ to its grade $f(I) = g$, where $I = (t_1, t_2, t_3, t_4, s)$ with $t_i, s \in \mathbb{N}_0$, $t_i, s \leq 100$ for $i \in \{1,2,3,4\}$ and $t_i < t_j$ for $i > j$ and $i,j \in \{1,2,3,4\}$ where $t_i$ are the thresholds and $s$ is the score achieved by a student and $g = \{A, B, C, D, F\}$ is the corresponding grade with the order $A \succ B \succ C \succ D \succ F$.    
\\\\
\noindent \textbf{Metamorphic Relations}: \textbf{\textit{1)}} As first relation, we require that a better numeric score leads to an equal or better discrete grade. For a given valid random input $I = (t_1, t_2, t_3, t_4, s)$ and the program $f$ we calculate $f(I) = g_I$. After that, we create a manipulated input $I' = (t_1, t_2, t_3, t_4, s + k)$ based on $I$ with $k \in \{0, \dots, 100-s\}$ and grade $f(I') = g_{I'}$ as follow-up test. The metamorphic relation holds if $g_{I'} \succeq g_I$. \textbf{\textit{2)}} Second, we define the opposite relation which requires that a lower numeric score leads to an equal or worse discrete grade. We create a manipulated input $I' = (t_1, t_2, t_3, t_4, s - k)$ based on $I$ with $k \in \{0, \dots, s\}$ and grade $f(I') = g_{I'}$ as follow-up test. This second metamorphic relation holds if $g_{I'} \preceq g_I$.
\\\\
\noindent \textbf{Small or Large Problem:}
\\\\
\noindent \textbf{Definition}: The generated program should classify a given integer either as small, large, or in between. More precise, we search for a function $f$ that maps a given integer $I = n \in \mathbb{Z}$ to its label $f(I) = l$, where $l = \{\texttt{"small"}, \texttt{""}, \texttt{"large"}\}$ with the order $\texttt{"small"} \prec \texttt{""} \prec \texttt{"large"}$. The function $f$ should return \texttt{"small"} if $n < 1,000$, \texttt{"large"} if $n \geq 2,000$, and an empty string (\texttt{""}) otherwise.
\\\\ 
\noindent \textbf{Metamorphic Relations}: \textbf{\textit{1)}} First, we require that for an increased input the label also has to stay equal or increase according to the defined label ordering. For a given integer $I = n$ we compute the resulting label $f(I) = l_I$. Following this, we manipulate the input $I' = n + k$ with a random $k \in \mathbb{N}$ with the corresponding label $f(I') = l_{I'}$. The relation holds if $l_{I'} \succeq l_I$. \textbf{\textit{2)}} Consequently, as second relation, we define that for an decreased input the resulting label has to stay equal or decrease according to the label ordering. For the given input $I = n$ we calculate the label $f(I) = l_I$. Then, we manipulate the input $I' = n - k$ with $k \in \mathbb{N}$ and $f(I') = l_{I'}$. If $l_{I'} \preceq l_I$, the relation is satisfied.

\subsection{Incorporating Metamorphic Testing in GP}

MTGP benefits from the use of lexicase selection, since lexicase considers the individual cases instead of an aggregated fitness value \cite{spector2012assessment}. Consequently, different types of (training) cases can be used simultaneously for the selection. In MTGP, these are hand-labeled training cases for which we know the input and the corresponding output, as well as cases based on metamorphic relations.

For the hand-labeled training cases, both the inputs and the outputs are known. In order to check whether a candidate program solves a training case or not, we run the program with the given input $I$ and check whether the generated output $f(I)$ matches the expected (in a real-world scenario hand-labeled) output $O$. If $f(I) = O$, then the test is passed, otherwise not.

To test the metamorphic relations defined for a benchmark problem, we provide random inputs as base inputs for which the outputs do not need to be known. Thus, any number of random entries can be generated at low cost. In our experiments, we have ensured that these random inputs do not already exist in the training or test set. In practice, they could be chosen freely. To check whether a candidate program satisfies a metamorphic relation or not, we execute it with a given base input $I$ (one of the randomly generated inputs) and save the output $f(I)$. We then change the input randomly according to the manipulation rule of the metamorphic relation to create the follow-up input $I'$ and run the program again. The test is passed if the observed outputs $f(I)$ and $f(I')$ satisfy the metamorphic relation, otherwise it is failed. Since we have defined two metamorphic relations for all benchmark problems considered in the experiments, we determine at random, with probability $p = 0.5$, which of the two relations is used each time a base input is selected.

One of the key advantages of using metamorphic testing this way is that during a GP run always many different metamorphic tests are executed, since the manipulation of the given basic input happens randomly. Our hope is, that this further supports the generalizability of the generated programs.

\section{Experiments and Discussion}\label{sec:experiments}

In this section, we study the performance of MTGP and analyze how well the generated programs generalize to previously unseen tests cases. 
To analyze how the generalization ability depends on the number of given training cases, we perform experiments for different labeled training set sizes. Below, we describe the experimental setup and discuss our results. 

\subsection{Experimental Setup}

For the implementation of MTGP, we use the PonyGE2 framework \cite{fenton2017ponyge2}. Our used grammars are based on the program synthesis grammars provided by the PonyGE2 framework for the problems from the benchmark suite \cite{helmuth2015general} and follow the principle suggested by Forstenlechner et al. \cite{forstenlechner2017grammar} according to which the grammar of a problem is restricted to the data types (and dependent functions) that are defined for the input and output of the considered problem, in addition to required base data types (e.g., like \texttt{integer} and \texttt{Boolean}). This allows to keep the used grammars small and effective but still expressive.

We initialize a run with position independent grow \cite{fagan2016exploring} and use a population size of $1,000$. The maximum initial tree depth is set to $10$ and the maximum overall tree depth is limited to $17$. As variation operators, we use sub-tree crossover and sub-tree mutation. For the sub-tree crossover we set the probability to $0.9$ and for the sub-tree mutation we set the number of mutation steps to $1$. As mentioned above, we use lexicase \cite{spector2012assessment} as selection method. A GP run is stopped either after a program is found that solves all labeled training cases and all defined metamorphic tests or after $300$ generations. 

For every considered benchmark problem, we have $200$ labeled training cases that we can use in the experiments. The test set consisting of $1,000$ labeled cases is used to check if a candidate program also generalizes to unseen cases. Further we provide a large set of randomly generated inputs ($800$ for each benchmark problem) which we use as base inputs for checking if the defined metamorphic relations hold for a candidate program or not. In the experiments, we choose from these available training cases and randomly generated inputs depending on the considered configuration as specified below.

\subsection{Results and Discussion}

As we investigate the impact of using metamorphic testing in GP, we compare a standard GP approach, which only uses the training data, and the novel MTGP approach, which also includes metamorphic tests. In addition, we examine how standard GP and MTGP perform on different training set sizes $|T_{\mathrm{training}}|$.
\definecolor{lightgray}{gray}{0.9}
\begin{table}
\centering
\caption{Success rates on the training set $s_{\mathrm{training}}$ and the test set $s_{\mathrm{test}}$ achieved by standard GP and MTGP for all studied labeled training set sizes $|T_{\mathrm{training}}|$ and benchmark problems. For MTGP, we use in addition $200 - |T_{\mathrm{training}}|$ metamorphic tests. Best success rates achieved on the test set $s_{\mathrm{test}}$ are printed in \textbf{bold} font.}
\label{tab:resultssuccessrates}
\renewcommand{\arraystretch}{1.3}
\begin{tabular}{lccccc}
\hline
 &  & \multicolumn{2}{c}{\textbf{Standard GP}} & \multicolumn{2}{c}{\textbf{MTGP}} \\
\textbf{Problem}\;\;\;\;\;\;\;\;\;\;\;\;\;\; & \;\;\;\;\boldsymbol{$|T_{\mathrm{training}}|$}\;\;\;\; & \;\;\;\;\boldsymbol{$s_{\mathrm{training}}$}\;\;\;\; & \;\;\;\;\boldsymbol{$s_{\mathrm{test}}$}\;\;\;\; & \;\;\;\;\boldsymbol{$s_{\mathrm{training}}$}\;\;\;\; & \;\;\;\;\boldsymbol{$s_{\mathrm{test}}$}\;\;\;\;\\
\hline
\hline 
Count Odds &  25 &                43 &            \textbf{28} &           32 &            20 \\
\rowcolor{lightgray}
           &  50 &                59 &            \textbf{46} &           43 &            35 \\ 
           & 100 &                70 &            \textbf{64} &           61 &            59 \\  
\rowcolor{lightgray}
           & 200 &                83 &            81 &            - &             - \\
\hline
Grade     &   25 &                83 &             4 &           74 &            \textbf{14} \\ 
\rowcolor{lightgray}
          &   50 &                86 &            12 &           76 &            \textbf{15} \\  
          &  100 &                81 &            23 &           84 &            \textbf{33} \\  
\rowcolor{lightgray}
          &  200 &                93 &            45 &            - &             - \\
\hline
Small or Large &   25 &                94 &             2 &             58 &             \textbf{7} \\   
\rowcolor{lightgray}
               &   50 &                79 &            11 &             77 &            \textbf{25} \\  
               &  100 &                91 &            \textbf{35} &             86 &            29 \\  
\rowcolor{lightgray}
               &  200 &                89 &            43 &              - &             - \\
\hline
\end{tabular}
\end{table}
Therefore,  we analyze for both approaches the labeled training set sizes $25$, $50$, $100$, and $200$. MTGP also uses $200 - |T_{\mathrm{training}}|$ metamorphic tests so that MTGP considers exactly $200$ tests during the training phase (e.g., for $|T_{\mathrm{training}}| = 25$, MTGP uses $175$ metamorphic tests).

As a first step, we study the achieved success rates on the training set $s_{\mathrm{training}}$ and the test set $s_{\mathrm{test}}$ as a performance indicator. The success rate on the training set $s_{\mathrm{training}}$ measures the percentage of runs in which a program is found that is successful on all given training cases (including metamorphic tests for MTGP). To determine the success rate on the test set 
$s_{\mathrm{test}}$, we take from each run the candidate program that performs best on the training data and measure the percentage of candidate programs that solve all previously unseen test cases. For every studied configuration, we performed $100$ runs.

Table~\ref{tab:resultssuccessrates} shows the success rates on the training set $s_{\mathrm{training}}$ and the test set $s_{\mathrm{test}}$ achieved by standard GP and MTGP for all studied labeled training set sizes $|T_{\mathrm{training}}|$ (and corresponding metamorphic tests) and benchmark problems. Best success rates achieved on the test set $s_{\mathrm{test}}$ are printed in \textbf{bold} font.

For most configurations, we see that standard GP achieves a higher success rate on the training data $s_{\mathrm{training}}$ compared to MTGP. Our assumption is that this is because the additional metamorphic tests prevent an overfitting to the training data. On the test data, MTGP performs best for the Grade problem and for most of the configurations of the Small or Large problem (compared to the corresponding standard GP runs). For the Count Odds problem best results for $s_{\mathrm{test}}$ are achieved with standard GP. 

\begin{table}
\centering
\caption{Generalization rate $G$ achieved by standard GP and MTGP for all studied labeled training set sizes $|T_{\mathrm{training}}|$ and benchmark problems. For MTGP, we use $200 - |T_{\mathrm{training}}|$ metamorphic tests in addition to the considered labeled training cases. Best results are printed in \textbf{bold} font.}
\label{tab:resultsgen1}
\renewcommand{\arraystretch}{1.3}
\begin{tabular}{lccccc}
\hline
 &  & \multicolumn{2}{c}{\textbf{Generalization rate} \boldsymbol{$G$}} \\
\textbf{Problem}\;\;\;\;\;\;\;\;\;\;\;\;\;\; & \;\;\;\;\;\;\boldsymbol{$|T_{\mathrm{training}}|$}\;\;\;\;\;\; & \;\;\;\;\;\;\;\textbf{Standard GP}\;\;\;\;\;\;\; & \;\;\;\;\;\;\;\textbf{MTGP}\;\;\;\;\;\;\;\\
\hline
\hline 
Count Odds     &  25 &  \textbf{65.116} & 62.5 \\
\rowcolor{lightgray}
               &  50 &  77.966 & \textbf{81.395} \\
               & 100 &  91.429 & \textbf{96.721} \\
\rowcolor{lightgray}
               & 200 &  97.59 & - \\
\hline 
Grade          &  25 &   4.819 & \textbf{18.919} \\
\rowcolor{lightgray}
               &  50 &  13.953 & \textbf{19.737} \\
               & 100 &  28.395 & \textbf{39.286} \\
\rowcolor{lightgray}
               & 200 &  48.387 & - \\
\hline 
Small or Large &  25 &  2.128 & \textbf{12.069} \\
\rowcolor{lightgray}
               &  50 & 13.924 & \textbf{32.468} \\
               & 100 & \textbf{38.462} & 33.721 \\
\rowcolor{lightgray}
               & 200 & 48.315 & - \\
\hline
\end{tabular}
\end{table}

More important than the pure success rates, however, is how well the programs found on the training data generalize to unseen test cases. In practice, a program synthesis approach which is known for its high generalization can simply be executed again if no solution was found in the first run. If the generalization is expected to be poor, it is unclear whether solutions found on the training data also work on previously unseen test cases, regardless of the success rate achieved on the training data. Large sets of additional test cases (to check for generalizability) are not available in a real-world scenario as manually labeling additional input/output examples is far too expensive. Therefore, in the second step, we analyze the generalization rate 
\begin{displaymath}
G = \frac{s_{\mathrm{test}}}{s_{\mathrm{training}}}\cdot 100.
\end{displaymath}

Table~\ref{tab:resultsgen1} shows the generalization rate $G$ achieved by standard GP and MTGP for all studied labeled training set sizes $|T_{\mathrm{training}}|$ (and corresponding metamorphic tests) and benchmark problems. Again, best generalization rates $G$ are printed in \textbf{bold} font.

We see that on average, best generalization rates $G$ are achieved with MTGP, as MTGP performed best in $7$ out of $9$ configurations where we have results for standard GP as well as for MTGP. The differences are particularly obvious for the Grade and the Small or Large problem when only a small labeled training set ($|T_{\mathrm{training}}| = 25$) is used. MTGP achieves for the Grade problem a generalization rate $G$ of $18.919$ and for the Small or Large problem of $12.069$ while standard GP achieves only $4.819$ and $2.128$, respectively. 

\begin{table}[b!]
\centering
\caption{Generalization rate $G$ achieved by standard GP and MTGP for all considered benchmark problems. All configurations use as labeled training set size $|T_{\mathrm{training}}| = 200$. For MTGP, we use $800$ metamorphic tests in addition to the considered labeled training cases. Best results are printed in \textbf{bold} font.}
\label{tab:resultsgen2}
\renewcommand{\arraystretch}{1.3}
\begin{tabular}{lccccc}
\hline
 &  & \multicolumn{2}{c}{\textbf{Generalization rate} \boldsymbol{$G$}} \\
\textbf{Problem}\;\;\;\;\;\;\;\;\;\;\;\;\;\; & \;\;\;\;\;\;\boldsymbol{$|T_{\mathrm{training}}|$}\;\;\;\;\;\; & \;\;\;\;\;\;\;\textbf{Standard GP}\;\;\;\;\;\;\; & \;\;\;\;\;\;\;\textbf{MTGP}\;\;\;\;\;\;\;\\
\hline
\hline 
Count Odds      & 200 & 97.59 & \textbf{100.0} \\
\rowcolor{lightgray}
Grade           & 200 & 48.387 & \textbf{61.538} \\
Small or Large  & 200 & 48.315 & \textbf{62.069} \\
\hline
\end{tabular}
\end{table}

So far we have only studied MTGP with a reduced labeled training set. But can the generalization rate even be increased compared to standard GP even if the complete labeled training set ($|T_{\mathrm{training}}| = 200$) is used during the run? To answer this question, we run MTGP this time with $800$ metamorphic tests. 

Table~\ref{tab:resultsgen2} shows the achieved generalization rates $G$ for standard GP and MTGP for all considered benchmark problems. As before, best generalization rates $G$ are printed in \textbf{bold} font.

Wee see that even when the complete labeled training set $T_{\mathrm{training}}$ is used during a run, using metamorphic testing can improve the generalization rate $G$. MTGP performs best on all studied benchmark problems. For the Count Odds problem, we even achieve a perfect generalization rate $G$ of $100$.

In summary, the generalization ability of the generated programs can be increased by using metamorphic testing. On average, best generalization rates are achieved with MTGP. The major advantage of metamorphic tests is that they do not require the expensive manual calculation of the expected outputs, since they work exclusively with random inputs which can be generated automatically.

\section{Conclusion}\label{sec:conclusion}

GP \cite{cramer1985representation,koza1992genetic} is an evolutionary approach that is well known for its performance in automatic program synthesis. Even if GP is competitive in performance to other state-of-the-art program synthesis approaches \cite{sobania2022choose}, it is not yet mature enough for a practical use in real-world software development, as many input/output examples are usually required during the training process to generate programs that also generalize to unseen test cases. As in practice, the training cases have to be labeled manually by the user which is very expensive, we need a supplementary approach to check the program behavior with a lower number of manually defined training cases.

With metamorphic testing~\cite{chen1998metamorphic}, we do not need labeled input/output examples. The program is executed multiple times, first on a given input (which can be generated randomly) and followed by related inputs to check whether certain user-defined metamorphic relations hold between the observed outputs. 

Therefore, in this work we suggested MTGP, an approach that combines metamorphic testing and GP and studied its performance and the generalization ability of the generated programs on common program synthesis benchmark problems. Further, we analyzed how the generalization ability depends on the number of given training cases and performed experiments for different labeled training set sizes.

We found that incorporating metamorphic testing in combination with hand-labeled training cases leads to a higher generalization rate than the exclusive use of hand-labeled training cases in almost all configurations studied in our experiments, including those using smaller labeled training sets as usual in GP-based program synthesis. Even with the largest considered labeled training set, the generalization rate could be increased by a large margin on all studied benchmark problems with the use of metamorphic testing. Consequently, we recommend researchers to use metamorphic testing in their GP approaches if the labeling of the training data is an expensive process in the considered application domain.  

In future work, we will study MTGP on additional program synthesis benchmark problems and further analyze the usage of the metamorphic tests as well as the given labeled training cases during a run to gain a deeper understanding of the implications of incorporating metamorphic testing in GP.

\bibliographystyle{splncs04}
\bibliography{mybibliography}
\end{document}